\documentclass[11pt]{article}

\usepackage{fullpage,latexsym,amsmath,amsfonts,graphicx,hyperref}

\usepackage{qcircuit}

\hypersetup{
    colorlinks,
    linkcolor={blue!100!black},
    citecolor={blue!100!black},
}

\usepackage{xcolor}
\def\01{\{0,1\}}
\newcommand{\ceil}[1]{\lceil{#1}\rceil}

\newcommand{\eps}{\varepsilon}
\newcommand{\ket}[1]{|#1\rangle}
\newcommand{\bra}[1]{\langle#1|}

\newcommand{\braket}[2]{\langle#1|#2\rangle} 
\newcommand{\Tr}{\mbox{\rm Tr}}

\newcommand{\E}{\mathbb{E}}

\newtheorem{theorem}{Theorem}

\newcommand{\errorprob}{\kappa}

\begin{document}

\title{Adaptive quantum phase estimation can be better than non-adaptive}
\author{Noah Linden\thanks{School of Mathematics, University of Bristol. Partially supported by the UK Engineering and Physical Sciences Research Council through grant EP/T001062/1. {\tt n.linden@bristol.ac.uk}}
\and
Ronald de Wolf\thanks{QuSoft, CWI and University of Amsterdam, the Netherlands. Partially supported by the Dutch Research Council (NWO) through Gravitation-grant Quantum Software Consortium, 024.003.037. {\tt rdewolf@cwi.nl}}
}
\maketitle

\begin{abstract}
Quantum phase estimation is one of the most important tools in quantum algorithms. It can be made non-adaptive (meaning all applications of the unitary $U_\phi$ happen simultaneously) without using more applications of $U_\phi$, albeit at the expense of using many more qubits. It is also known that there is no advantage for adaptive algorithms in the case where the phase that needs to be estimated is arbitrary or is uniformly random. Here we give examples of a special case of phase estimation, with a promise on the values that the unknown phase can take, where adaptive methods \emph{are} provably better than non-adaptive methods by a factor of nearly~2 in the number of uses of $U_\phi$. 
We also prove some upper bounds on the maximum advantage that adaptive algorithms for phase estimation can achieve over non-adaptive ones.
\end{abstract}

\section{Introduction}

\subsection{Phase estimation}

In phase estimation we have the ability to apply a unitary $U_\phi$ (possibly controlled by another qubit) and are given a copy of an eigenstate $\ket{\Psi}$ of $U_\phi$ with eigenvalue $e^{2\pi i\phi}$.
The task is to estimate the phase $\phi\in[0,1)$.
Kitaev~\cite{kitaev:stabilizer} showed how to estimate $\phi$ with additive error $\delta$ (modulo~1) using roughly $1/\delta$ applications of $U_\phi$.
The standard circuit for this is as follows. We use a set of $k=\ceil{\log(1/\delta)}$ auxiliary qubits starting in state $\ket{0}$ which will end up containing a $k$-bit estimate of $\phi$.
These $k$ qubits are first put into a uniform superposition over all $j=0,\ldots,2^k-1$, which controls how many times $U_\phi$ is applied to $\ket{\Psi}$.
Because $\ket{j}\otimes U_\phi^j\ket{\Psi}=e^{2\pi i j\phi}\ket{j}\otimes\ket{\Psi}$, we obtain the following $k$-qubit state (tensored with our one copy of $\ket{\Psi}$)
\begin{equation}
    \frac{1}{\sqrt{2^k}}\sum_{j=0}^{2^k-1}e^{2\pi ij\phi}\ket{j}.
\end{equation}
By writing the binary expansion of $j$ as a sum of powers of 2, this can be nicely factored as
\begin{equation}\label{eq:PEproductstate}
    \frac{1}{\sqrt{2}}(\ket{0}+e^{2\pi i2^{k-1}\phi}\ket{1})\otimes
    \frac{1}{\sqrt{2}}(\ket{0}+e^{2\pi i2^{k-2}\phi}\ket{1})\otimes\cdots\otimes\frac{1}{\sqrt{2}}(\ket{0}+e^{2\pi i\phi}\ket{1}).
\end{equation}
One can think of this as applying the phase $\phi$ $2^{k-1}$ times on the first qubit, $2^{k-2}$ times on the second qubit, etc.
An inverse quantum Fourier transform (QFT, denoted $Q$ below) will then convert this state into a state whose amplitudes are concentrated around the best $k$-bit approximations of $\phi$ (mod~1).
In fact, if $\phi$ can be written exactly with $k$ bits of precision, then the circuit produces $\phi$ with probability~1, because the state of Eq.~\eqref{eq:PEproductstate} is then $Q\ket{2^k\phi}$.
The circuit is depicted below.

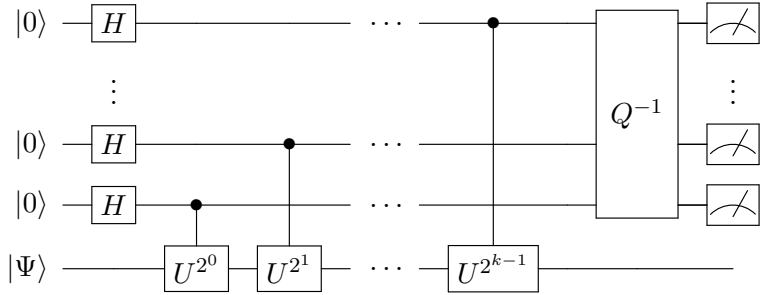
\begin{figure}[htb]
     \centering
\begin{equation*}
    \Qcircuit @C=1em @R=.7em {
\lstick{\ket{0}} & \gate{H} & \qw & \qw & \qw & \push{\!\!\dots\;\;} & \ctrl{4} & \qw & \multigate{3}{Q^{-1}} & \meter \qw\\
& \push{\vdots} & & & & & & & & \vdots \\
\lstick{\ket{0}} & \gate{H} & \qw & \ctrl{2} & \qw & \push{\!\!\dots\;\;} & \qw & \qw & \ghost{Q^{-1}} & \meter\qw \\
\lstick{\ket{0}} & \gate{H} & \ctrl{1} & \qw & \qw & \push{\!\!\dots\;\;} & \qw & \qw & \ghost{Q^{-1}} & \meter\qw \\
\lstick{\ket{\Psi}} & \qw & \gate{U^{2^0}} & \gate{U^{2^1}} & \qw & \push{\!\!\dots\;\;} & \gate{U^{2^{k-1}}} & \qw & \qw & \qw\\
}
\end{equation*}
     \caption{The standard circuit for phase estimation}
     \label{fig:PE}
 \end{figure}
 
Many quantum algorithms can be cast as an application of phase estimation~\cite{cemm:revisited}, including integer factoring~\cite{shor:factoring,kitaev:stabilizer} and amplitude estimation~\cite{bhmt:countingj} (used for approximate counting).
Phase estimation is a crucial subroutine in quantum walks (used there to reflect about the uniform superposition over all edges of a graph~\cite{mnrs:searchwalk}), in the negative-weights adversary bound in query complexity (used there to turn a dual-SDP solution into a quantum algorithm~\cite{reichardt:tight,lmrss:stateconv}) and in many applications in chemistry and physics (used there for instance to measure the energy of a state).

A fundamental question for quantum computing is to what extent quantum algorithms can be \emph{parallelized}. 
Parallelization reduces the time qubits need to stay coherent, at the expense of having to do many operations simultaneously, and typically using more auxiliary qubits.\footnote{Thus in practical applications, low-depth algorithms can have reduced impact of noise because there is less time for the qubits to decohere.} We call a phase estimation algorithm \emph{non-adaptive} if all its applications of $U_\phi$ are applied in parallel, meaning at the same time step.
This corresponds to full parallelization, at least of the applications of~$U_\phi$.
In contrast, in an \emph{adaptive} algorithm earlier applications of $U_\phi$ can affect the state on which later $U_\phi$'s are applied. 
For example, the standard circuit for phase estimation (Figure~\ref{fig:PE}) is adaptive.
The goal of this paper is to describe cases where adaptive phase estimation algorithms can be provably more efficient than non-adaptive ones.\footnote{Our use of the term ``non-adaptive'' follows computer-science terminology, and should not be confused with other uses in quantum sensing~\cite[Section~VIII.A.2]{DRP:quantumsensing} and quantum information~\cite{DFLS:adaptive}.} 

We note that our notion of an adaptive algorithm allows for measurements which are performed in the middle of the computation, between applications of $U_\phi$, and subsequent actions of the algorithm are dependent on the classical outcomes of those measurements.
However, if we only care about the number of applications of~$U_\phi$, then by the so-called ``principle of deferred measurement'' (see e.g.~\cite[Section~4.4]{nielsen&chuang:qc}) we can move all measurements towards the end, making everything coherent without increasing the number of uses of~$U_\phi$. This works by replacing a measurement of a qubit by a CNOT gate that ``copies'' the qubit to an auxiliary qubit that is initially $\ket{0}$ and that is not touched further in the computation.
Later parts of the computation are then conditioned on the unmeasured qubit rather than on a classical binary measurement outcome.
It is not hard to show that the final output distribution, after tracing out the auxiliary qubits, will be exactly the same as for the original algorithm that used intermediate measurements.
The classical control that might  select the next operation to do based on intermediate measurement outcomes can be simulated with a lot of controlled operations. This can significantly increase the total number of gates in the circuit, but not the number of applications of (controlled) $U_\phi$.

\subsection{A basic problem in sensing}\label{ssec:basicsensing}

The following is a basic problem in \emph{quantum sensing}: estimate the parameter $\phi\in[0,1)$ given the ability to apply a one-qubit unitary
\begin{equation}\label{eq:basicU}
U_\phi=\left(\begin{array}{cc}1 & 0\\
0 & e^{2\pi i\phi}\end{array}\right),
\end{equation}
This is of course just the special case of phase estimation where the eigenstate $\ket{\Psi}$ is $\ket{1}$, and $\ket{0}$ is an eigenstate
with eigenvalue~1 (equivalently, eigenphase~0).
Note that if we can implement such a $U_\phi$ then we can also implement controlled-$U_\phi$ ourselves (by swapping the target with the eigenvalue-1 state $\ket{0}$ if the control bit is 0), so we no longer need to \emph{assume} the ability to apply controlled version of~$U_\phi$.
One might also hope that knowing the eigenvectors could make phase estimation cheaper, but that is not the case:
the standard circuit (with $\ket{\Psi}=\ket{1}$) is still optimal for obtaining a good $k$-bit approximator of $\phi$ with high success probability, both in the worst-case and in the average-case where $\phi$ is uniformly random.%
\footnote{This optimality for the average case of uniformly random $\phi\in[0,1)$  was proved by van Dam et al.~\cite{DDMM:optimalPE}. It is not difficult to see that the same (average-case) proof will work for $\phi$ being a uniformly random integer multiple of $2^{-k}$. In fact, the worst case for the latter problem can be reduced to the average case by changing the $U_\phi$ by adding a uniformly random offset in $\{0,2^{-k},2\cdot2^{-k},3\cdot 2^{-k},\ldots, 1-2^{-k}\}$ to $\phi$ before the algorithm (and subtracting it from the obtained estimate at the end), as is done in~\cite{linden&wolf:verifyQFT}.
This shows that the average-case phase estimation problem is no easier than the worst-case for this set of phases.
\cite{DDMM:optimalPE} considered a cost function $C(\phi,\tilde{\phi})$ between the actual unknown phase $\phi$ and the algorithm's estimator $\tilde{\phi}$, under the assumption that $C(\cdot,\cdot)$ only depends on the difference $\phi-\tilde{\phi}$. They proved their result by showing how every adaptive algorithm can be massaged into a non-adaptive one with the same number of applications of~$U_\phi$ and the same expected cost function (i.e., the expectation $\E[C(\phi,\tilde{\phi})]$ does not go up, where the expectation is taken over a prior distribution over $\phi$ and over the algorithm's randomness in producing an estimator $\tilde{\phi}$ on input $\phi$). In our paper the cost measure is the ``01-loss'' $C(\phi,\tilde{\phi})=1-\delta_{\phi,\tilde{\phi}}$, which satisfies their assumption. However, their result does not contradict our factor-2 improvement because their proof crucially only preserves the expectation of $C(\phi,\tilde{\phi})$ under a \emph{uniform} prior distribution over all $\phi\in[0,1)$, while our phases~$\phi$ come from a very specific phase-set~$F$.}

In this simple case, phase estimation can be made non-adaptive without using more applications of $U_\phi$ than the standard circuit.
The idea is to produce each one of the $k$ qubits of the product state of Eq.~\eqref{eq:PEproductstate} using its own ``cat state''. This is similar to a well known  approach in quantum sensing, see for instance this comprehensive 
review~\cite[Sec.~VIII.A \&\ IX.C]{DRP:quantumsensing}. For each $p=2^{k-1},2^{k-2},\ldots,2,1$ we create the $p$-qubit state
\[  
\frac{1}{\sqrt{2}}(\ket{0^p}+\ket{1^p})
\]
starting from $\ket{0^p}$ using a Hadamard and $p-1$ CNOTs. We can then apply $U_\phi^{\otimes p}$ to induce the phase $e^{2\pi i p\phi}$ on $\ket{1^p}$, and invert the CNOTs to end with the qubit in the state $\frac{1}{\sqrt{2}}(\ket{0}+e^{2\pi i p\phi}\ket{1})$. Doing this in parallel for each of the $k$ qubits in the state \eqref{eq:PEproductstate}, we prepare that state in a way where the applications of $U_\phi$ are all in parallel, all at the same timestep, thus ``flattening'' the circuit without using more applications of $U_\phi$ than the standard adaptive circuit.\footnote{Something similar would work to ``flatten'' arbitrary phase estimation, with arbitrary multi-qubit $U_\phi$ and unknown $\ket{\Psi}$, but it would require us to prepare ket-states of the form $\frac{1}{\sqrt{2}}(\ket{\Psi_1}^{\otimes p}+\ket{\Psi_2}^{\otimes p})$ where $\ket{\Psi_1}$ and $\ket{\Psi_2}$ are eigenstates of $U_\phi$ whose relative phase is $\phi$. This seems much more demanding than a one-qubit unitary~$U_\phi$ as in Eq.~\eqref{eq:basicU}.} The price we pay for this parallelization is the large number $O(2^k)$ of qubits needed to prepare the cat states. This is an extreme example of a width-vs-depth tradeoff, where we greatly reduce the depth of the circuit at the expense of greatly increasing the number of qubits. Note that the circuits for preparing the cat states and for the final inverse QFT, are not constant-depth, but can be implemented in logarithmic depth; for cat states this can be done by a binary tree of CNOT gates, and for the QFT this was done by Cleve and Watrous~\cite{cleve&watrous:parfourier}.

\subsection{Our result: adaptive phase-estimation algorithms can be twice as efficient}\label{subsection:our-result}

In many applications of phase estimation and sensing, the phase $\phi$ that we want to estimate is not completely arbitrary, but comes from some limited set $F$ of possible phases. For example, it could be that the possible phases correspond to  certain specific energies, like the specific spectral lines of some molecule.
We want to find examples in this setting where adaptive phase estimation is provably more efficient than non-adaptive phase estimation, meaning parallelization comes at a  cost in the number of applications of $U_\phi$ (in contrast to promiseless phase estimation, where the standard circuit can be parallelized without using more applications of $U_\phi$, as we argued above).

We start in Section~\ref{sec:4phases} with a simple example with 4 possible values for $\phi$: $0,2^{-k},1/2,1/2+2^{-k+1}$.
Note that the first two phases can easily be distinguished from the other two phases by looking at $U_\phi\ket{+}$, which is equal or close to $\ket{+}$ for the first two phases, and equal or close to $\ket{-}$ for the other two phases. An adaptive algorithm can make this distinction first, with error probability $O(2^{-k})$.
If it finds itself in the first case, then it uses $2^{k-1}$ applications of $U_\phi$ to prepare the qubit $U_\phi^{2^{k-1}}\ket{+}$ to resolve the difference between phases $0$ and $2^{-k}$. In the second case it uses $2^{k-2}$ applications of $U_\phi$ to prepare the qubit $U_\phi^{2^{k-2}}\ket{+}$ to resolve the difference between $1/2$ and $1/2+2^{-k+1}$. So in the worst case this uses $\frac{1}{2}2^k+1$ applications of $U_\phi$ to learn which of the 4 possibilities $\phi$ is.

This upper bound of $\frac{1}{2}2^k+1$ applications of $U_\phi$ is optimal (up to lower-order terms) for distinguishing the 4 phases in $F$, even adaptively.
In fact  $(\frac{1}{2}-o(1))2^k$ applications of $U_\phi$ are already needed
just to distinguish the phases 0 and $2^{-k}$ with success probability $1-o(1)$ (where $o(1)$ denotes any function that goes to 0 with $k$), as follows. Up to a global phase and a change of basis, our $U_\phi$ can be thought of as a rotation matrix over angle $\pi\phi$.  By the principle of deferred measurement, we may assume w.l.o.g.\ that the distinguishing circuit is unitary, i.e., no intermediate measurements (this has a cost in extra auxiliary qubits and controlled operations, but not more applications of $U_\phi$).
If we want to distinguish the cases $\phi=0$ from $\phi=2^{-k}$, then the initial state needs to be rotated to two different states at angle $\pi/2-o(1)$, depending on which of the two cases we are in.  Since each application of $U_\phi$ can increase the angle between the states by at most $\pi\cdot 2^{-k}$ (see the appendix), we need at least $(\pi/2-o(1))/(\pi\cdot 2^{-k})=(\frac{1}{2}-o(1))2^k$ applications of $U_\phi$.\footnote{We focus on error probability $1-o(1)$ since this allows us to get precise constants (up to additive $o(1)$) in the number of uses of $U_\phi$. If, in contrast, we had wanted only constant error probability, say $0.01$, then $N=0.47 \cdot 2^k$ applications of $U_\phi$ would already suffice to distinguish phases 0 and $2^{-k}$, by preparing the qubit $U_\phi^N\ket{+}=\frac{1}{\sqrt{2}}(\ket{0}+\chi\ket{1})$, where $\chi=1$ if $\phi=0$ and $\chi=e^{2\pi i\, 0.47}\approx -0.982+ i0.187$ if $\phi=2^{-k}$. Doing a Hadamard gate and measuring in the computational basis gives outcome~0 with certainty if $\phi=0$, and gives outcome~1 with probability $\approx|(1.982-i0.187)/2|^2\approx 0.991$ if $\phi=2^{-k}$. } 

Hence our algorithm is essentially optimal for this 4-phase task.
This algorithm is adaptive: it first (cheaply) distinguishes the first pair of phases from the second pair, and then based on that outcome determines the next phase estimation it will perform.

How well can a \emph{non-adaptive} algorithm do for distinguishing these 4 phases?
Initially one might expect that such an algorithm has to separately do the work for recovering the $k$th and the $(k-1)$th bits of $\phi$, since it cannot adaptively find out first which of these two is needed, and the qubit $U_\phi^{2^{k-1}}\ket{+}$ that is used to learn the $k$th bit of $\phi$ gives no information about the $(k-1)$th bit of $\phi$. The number of applications of $U_\phi$ would then be $2^{k-1}+2^{k-2}=\frac{3}{4}2^k$.
However, we will show here that in fact $(\frac{2}{3}\pm o(1))2^k$ applications of $U_\phi$ are necessary and sufficient for non-adaptive phase estimation with error probability $1-o(1)$.
This is better than the naive non-adaptive algorithm, but still a factor 4/3 worse than the number of applications of $U_\phi$ that suffice for an adaptive algorithm.

Our upper bound proof for the number of uses of $U_\phi$ in the non-adaptive case finds explicit amplitudes for the initial state on which the $U_\phi$'s are applied in parallel. Our lower bound proof massages the existence of a non-adaptive algorithm with $N$ applications of $U_\phi$ into a linear constraint problem. Using Farkas's lemma we can then show the \emph{in}feasibility of that problem: obtaining a lower bound on the number of applications of $U_\phi$ by exhibiting a feasible solution for the \emph{dual} problem.
This approach via duality to rule out a suitable initial state is (to the best of our knowledge) a novel proof method to show the non-existence of a good quantum algorithm.

In Section~\ref{sec:2ellphases} we extend the 4-phase example to larger sets of phases. In particular, by choosing a set $F_\ell$ that consists of $2^\ell$ pairs of phases, where the two elements of each pair are separated by a different multiple of $2^{-k}$ (e.g., $2^{-k},2\cdot 2^{-k},3\cdot 2^{-k},\ldots$), we construct an instance of phase estimation where roughly $\frac{1}{2}2^k$ applications of $U_\phi$ still suffice for an adaptive algorithm, while for non-adaptive algorithms, $\frac{2^\ell}{2^\ell +1}2^k$ applications of $U_\phi$ are necessary and sufficient. 
For large constant $\ell$, this gives an advantage of nearly a factor of 2 for adaptive algorithms over non-adaptive ones.

 As we discuss in Section~\ref{sec:bettersep?}, these separations of a factor of~2 are essentially best-possible for the type of sets $F$ that we consider here.

\subsection{Other examples of adaptive advantage}
We give in this paper a small adaptive-vs-non-adaptive advantage for phase estimation.
This advantage is one of many examples of adaptive advantage in quantum computing, and more broadly in other branches of computing as well. It is not surprising that for some computational problems parallelization is not possible without significant overhead, leaving us with an adaptive advantage, though this is not always easy to prove. As a famous example, the Euclidean algorithm is an efficient adaptive algorithm for computing the greatest common divisor of two integers in polynomial time, and it is a longstanding conjecture that this algorithm cannot be parallelized to logarithmic depth.

Other settings in quantum computing allow for much larger (exponential) adaptive advantage, for example quantum communication complexity, based on versions of ``pointer jumping'' functions. There the input is a tree of instances of some smaller communication complexity problem, each internal problem instance's solution points to one of its children. These pointers chart a length-$k$ path from the root of the tree to a particular leaf, which then contains the solution to the overall problem. For problem-trees of depth $k$, $k$-round interactive communication protocols can just adaptively follow the length-$k$ path to the correct leaf; on the other hand, one can show that one-way (or even $(k-1)$-round) quantum communication protocols need exponentially more communication~\cite{kntz:qinteractionj}. One can similarly (and much more easily) construct exponential adaptive separations in adaptive vs.\ non-adaptive quantum \emph{query} complexity. 

In adaptive vs.\ non-adaptive \emph{channel discrimination}, \cite{HHLW:adapchannel} gave a pair of entanglement-breaking channels that can be perfectly discriminated adaptively by two channel evaluations, but for which every non-adaptive strategy that evaluates the channel a finite number of times has some positive error probability. More recently, \cite{SHW:adaptivestrat} showed that for discriminating two classical-quantum channels, adaptive and non-adaptive strategies lead to the same error exponents in various asymptotic settings, while for discriminating fully quantum channels adaptivity can lead to a better exponent. In the non-asymptotic regime with asymptotically vanishing type I error probability, \cite{BDSW:parallelization} showed that adaptive channel discrimination strategies can be made non-adaptive at the expense of a small loss in the error exponent and a small increase in the number of uses of the channel.

\section{First example: a set $F$ with 4 possible phases}\label{sec:4phases}

\subsection{The problem and an upper bound of $\frac{1}{2}2^k$ for adaptive algorithms} 

In this section we make our first example precise.
We consider the problem of phase estimation with a single-qubit unitary $U_\phi$ as in Eq.~\eqref{eq:basicU}, with the phase $\phi$ coming from the set 
\begin{equation}
F=\{0,2^{-k},1/2,1/2+2^{-k+1}\}.
\end{equation}
As we already argued in the introduction, an \emph{adaptive} algorithm can identify $\phi$ with error probability $O(2^{-k})$ using $\frac{1}{2}2^k+1$ applications of $U_\phi$.

\subsection{An upper bound of $\frac{2}{3}2^k$ for non-adaptive algorithms}

Because the 4 phases in the set $F$ are either close to or equal to 0, or close to or equal to 1/2, we can learn the most significant bit of $\phi$ with only one application of $U_\phi$, by just preparing $U_\phi\ket{+}$ and measuring in the $\ket{\pm}$ basis.
If $\phi\in\{0,2^{-k}\}$ then we will almost certainly  (with error probability $\exp(-k)$) get outcome $\ket{+}$, and if 
$\phi\in\{1/2,1/2+2^{-k+1}\}$ then we will almost certainly get outcome $\ket{-}$.
We can also non-adaptively learn the $k$th and $(k-1)$th bits of $\phi$ with success probability~1, by preparing $U_\phi^{\frac{1}{2}2^k}\ket{+}$  and  
$U_\phi^{\frac{1}{4}2^k}\ket{+}$ and doing a joint measurement on those 2 qubits.
This non-adaptive algorithm uses $\frac{1}{2}2^k+\frac{1}{4}2^k+1=\frac{3}{4}2^k+1$ applications of $U_\phi$.

However we will now describe a more efficient non-adaptive algorithm which uses roughly $\frac{2}{3}2^k$ applications of $U_\phi$, and which identifies $\phi\in F$ with error probability $\exp(-k)$.
An arbitrary non-adaptive algorithm with $N$ applications of $U_\phi$ would prepare some initial state 
\begin{equation}\label{eq:initialstate}
\ket{\psi}=\sum_{x\in\01^N}\alpha_x\ket{x}\ket{\eta_x}
\end{equation}
of $q\geq N$ qubits, apply $U_\phi^{\otimes N}\otimes I_{2^{q-N}}$
(which adds phase $e^{2\pi i|x|\phi}$ to $\ket{x}$, where $|x|$ is the Hamming weight of $x$), and do a measurement with $|F|$ outcomes corresponding to the possible phases $\phi\in F$.

We will now massage such an arbitrary non-adaptive protocol into a form that is a bit easier to analyse.
For $j=0,\ldots,N$, define the orthonormal states 
$\ket{\chi_j}=\frac{1}{\beta_j}\sum_{x\in\01^N:|x|=j}\alpha_x\ket{x}\ket{\eta_x}$, where $\beta_j=\sqrt{\sum_{x\in\01^N:|x|=j}|\alpha_x|^2}$ is a normalizing factor. Then we can write the state before the measurement as
\begin{equation}
\sum_{j=0}^N\beta_j e^{2\pi ij\phi}\ket{\chi_j}  
\end{equation}
By doing a unitary that is independent of $\phi$, we may assume the state before the measurement is actually the $(N+1)$-dimensional state
\begin{equation}
\ket{\psi_\phi}=\sum_{j=0}^N\beta_j e^{2\pi i j\phi}\ket{j} .
\end{equation}
A non-adaptive algorithm that uses $U_\phi$ a total of $N$ times, is thus fully described by the unit vector of amplitudes $(\beta_0,\ldots,\beta_N)$.

We will now show how to choose the $\beta_j$'s to obtain a non-adaptive algorithm for learning phases $\phi\in F$ with roughly $\frac{2}{3}2^k$ applications of $U_\phi$.
Our non-adaptive algorithm will have two registers: one containing one qubit, and one containing an $(N+1)$-dimensional state ($\ceil{\log N}$ qubits). The total number of applications of $U_\phi$ will be $N+1$: we act with $U_\phi$ on the first qubit and with the map $\ket{j}\mapsto e^{2\pi ij \phi}\ket{j}$ on the second register (which corresponds to the action of $U_\phi^{\otimes N}\otimes I$ on the original initial state of Eq.~\eqref{eq:initialstate}) to produce the final two-register state $\ket{\psi_\phi}$. We want the four states
$\ket{\psi_0},\ket{\psi_{2^{-k}}},\ket{\psi_{1/2}},\ket{\psi_{1/2+2^{-k+1}}}$ to be almost orthogonal, so we can distinguish them with high success probability using some specific measurement.  Specifically, we choose $N$ to be a multiple of 4 closest to $\frac 2 3 2^k$, thus $N=\frac 2 3 2^k+\Delta$, where $|\Delta|<2$, and we choose the initial state $\ket\psi$ to be
\begin{equation}\label{eq:stateforell=1}
\ket\psi = \ket +  \left( \frac{1}{\sqrt{3}}\ket{0}+    \frac{1}{\sqrt{3}}\ket{N/2}
+\frac{1}{\sqrt{3}}\ket{N}\right).
\end{equation}
Because we chose $N$ a multiple of 4, the most-significant bit of $\phi$ (i.e., whether $\phi$ is close to 0 or to 1/2) is irrelevant if we apply the phase $0$ or $N/2$ or $N$ times.

Thus 
\begin{eqnarray}
    \ket{\psi_0} &=&  \ket +  \left( \frac{1}{\sqrt{3}}\ket{0}+    
    \frac{1}{\sqrt{3}}\ket{N/2} +\frac{1}{\sqrt{3}}\ket{N}\right)\nonumber\\
    \ket{\psi_{2^{-k}}} & = &  \ket {+'}  \left( \frac{1}{\sqrt{3}}\ket{0}+    
    \eta\frac{1}{\sqrt{3}}\ket{N/2} +\eta^2\frac{1}{\sqrt{3}}\ket{N}\right)\nonumber\\
    \ket{\psi_{1/2}} &=& \ket -  \left( \frac{1}{\sqrt{3}}\ket{0}+    
    \frac{1}{\sqrt{3}}\ket{N/2} +\frac{1}{\sqrt{3}}\ket{N}\right)\nonumber\\
\ket{\psi_{1/2+2^{-k+1}}} & = &  \ket {-'}  \left( \frac{1}{\sqrt{3}}\ket{0}+    
    \eta^2\frac{1}{\sqrt{3}}\ket{N/2} +\eta^4\frac{1}{\sqrt{3}}\ket{N}\right),
\end{eqnarray}
where 
\begin{equation}
    \eta = \exp 2\pi i \frac {N} {2^{k+1}}.
\end{equation}
The states $ \ket {+'}=\frac {1}{\sqrt 2}(\ket 0 + e^{2\pi i/2^k}\ket 1) $ and  $\ket {-'} =\frac {1}{\sqrt 2}(\ket 0 - e^{2\pi i/2^{k-1}}\ket 1) $  are exponentially (in $k$) close to $\ket {+} $ and  $\ket {-} $ respectively.  Since we chose $N=\frac 2 3 2^k+\Delta$ with $|\Delta|<2$, we see that $\eta$ is exponentially close to a third root of unity:
\begin{equation}
    \eta = \exp\left( 2\pi i \left(\frac{1}{3} + \frac{\Delta}{2^{k+1}}\right)\right).
\end{equation}
Thus, defining $\omega$ as the third root of unity $\omega=\exp \frac{2}{3}\pi i$, the states $\ket{\psi_0},\ket{\psi_{2^{-k}}},\ket{\psi_{1/2}},\ket{\psi_{1/2+2^{-k+1}}}$ are exponentially close to 
\begin{eqnarray}\label{eq:Fourier-type-states}
    & & \ket +  \left( \frac{1}{\sqrt{3}}\ket{0}+    
    \frac{1}{\sqrt{3}}\ket{N/2} +\frac{1}{\sqrt{3}}\ket{N}\right)\nonumber\\
    & & \ket {+}  \left( \frac{1}{\sqrt{3}}\ket{0}+    
    \omega\frac{1}{\sqrt{3}}\ket{N/2} +\omega^2\frac{1}{\sqrt{3}}\ket{N}\right)\nonumber\\
    & & \ket -  \left( \frac{1}{\sqrt{3}}\ket{0}+    
    \frac{1}{\sqrt{3}}\ket{N/2} +\frac{1}{\sqrt{3}}\ket{N}\right)\nonumber\\
& & \ket {-}  \left( \frac{1}{\sqrt{3}}\ket{0}+    
    \omega^2\frac{1}{\sqrt{3}}\ket{N/2} +\omega\frac{1}{\sqrt{3}}\ket{N}\right)\label{eq:4perfectstates},
\end{eqnarray}
respectively.
We can now measure the first qubit in the Hadamard basis and the second register with a projective measurement that distinguishes the three pairwise-orthogonal states of the second register.
If we had one of the 4 states of Eq.~\eqref{eq:4perfectstates}, then this measurement would give the correct outcome with probability~1.
Since the state we are actually measuring is exponentially (in $k$) close to one of these 4 states, we will get the correct outcome except with exponentially 
small probability.

\subsection{A lower bound of $\frac{2}{3}2^k$ for non-adaptive algorithms}\label{ssec:LBell=1}

We note that the adaptive algorithm in Section \ref{subsection:our-result} achieves error probability that goes to 0 with~$k$. For fair comparison, we require that the non-adaptive algorithm also has error probability $\errorprob=o(1)$ (meaning $\errorprob$ goes to 0 with $k$).  Here we will prove that every non-adaptive algorithm needs at least $(\frac{2}{3}-o(1)) 2^k$ applications of $U_\phi$ to achieve  $\errorprob=o(1)$.  Thus our algorithm in the previous section is essentially best possible. This proves a constant-factor advantage of adaptive over non-adaptive phase-estimation 
algorithms: at most $\frac{1}{2} 2^k+1$ vs.\ more than $(\frac{2}{3}-o(1))2^k$ applications of~$U_\phi$.

Recall from the previous section that a non-adaptive algorithm with $N$ applications of $U_\phi$, is fully described  by the $(N+1)$-dimensional final states
\begin{equation}
\ket{\psi_\phi}=\sum_{j=0}^N\beta_j e^{2\pi i j\phi}\ket{j} .
\end{equation}
A non-adaptive algorithm with (worst-case) error probability $\errorprob$
will (among other things) have to distinguish phase $\phi=2^{-k}$ from phase~$\phi'=0$, and also distinguish phase 
$\phi=1/2+2^{-k+1}$ from $\phi'=1/2$.
This implies that the two complex inner products
\begin{equation}\label{eq:2innerprods}
   \braket{\psi_0}{\psi_{2^{-k}}}=\sum_{j=0}^N\beta_j^2 e^{2\pi i j/2^k}
\mbox{ and }
\braket{\psi_{1/2}}{\psi_{1/2+2^{-k+1}}}=\sum_{j=0}^N\beta_j^2 e^{4\pi i j/2^k} 
\end{equation}
both have small magnitude: at most $O(\sqrt{\errorprob})$, since the probability to ``confuse'' two states corresponds to their squared inner product.
Translating to 4 real equations, there must exist small real numbers  $\eps_1,\eps_2,\eps_3,\eps_4$ (possibly negative) of magnitude $O(\sqrt{\errorprob})$ such that
\begin{eqnarray}\label{eq:linear-system-l-equals-1}
\sum_{j=0}^N \cos(2\pi j/2^k)\beta_j^2  & = & \eps_1\\
\sum_{j=0}^N \sin(2\pi j/2^k)\beta_j^2  & = & \eps_2\\
\sum_{j=0}^N \cos(4\pi j/2^k)\beta_j^2  & = & \eps_3\\
\sum_{j=0}^N \sin(4\pi j/2^k)\beta_j^2  & = & \eps_4.
\end{eqnarray}
Let $A$ be the  $5\times (N+1)$ real matrix whose first row is all-1s, and whose last 4 rows correspond to the sines and cosines in the above 4 equations, respectively. Let $E$ be the column vector $(1,\eps_1,\eps_2,\eps_3,\eps_4)^T$. If there is a non-adaptive algorithm, then there will exist a vector $b\in\mathbb{R}^{N+1}$ (corresponding to the $\beta_j^2$, which sum to~1)
satisfying 
\begin{equation}\label{eq:farkasprimal}
Ab=E\mbox{ and }b\geq 0.
\end{equation}
{\bf Farkas's lemma}~\cite{farkas:lemma} (which may be found in any text on linear programming or convex geometry) says that the system \eqref{eq:farkasprimal} is \emph{unsatisfiable} iff the following system over variable-vector $y$ is \emph{satisfiable}
\begin{equation}\label{eq:farkasdual}
    y^T A\geq 0\mbox{ and }y^T E<0.
\end{equation}
Geometrically, this lemma says that either the vector $E$ is in the positive cone spanned by the columns of $A$ (with $b$ being the vector of positive coefficients), or there is a hyperplane that separates $E$ from that positive cone (in the sense that the hyperplane's normal vector $y$ has positive inner product with each of the columns of $A$ but negative inner product with $E$).

Accordingly, we can use Farkas's lemma to lower bound the cost of non-adaptive algorithms by just exhibiting 5 numbers: if we can find a $y\in \mathbb{R}^5$ satisfying the latter system, then we have proved that every non-adaptive algorithm that can identify the phases in $F$ with error probability~$o(1)$, needs to use more than $N$ applications of $U_\phi$.

Our strategy for exhibiting such a 5-dimensional vector will be as follows. Firstly, we will find a setting of the $y_j$'s with $y_0=0$ that satisfies the first inequality in \eqref{eq:farkasdual}. But this choice does not necessarily satisfy the second inequality of~\eqref{eq:farkasdual}.  We then choose $y_0$ to be a small negative value, in order to satisfy the second inequality; but we will also add small amounts to the other $y_j$'s (depending on $y_0$) to ensure that the first inequality is still satisfied; we will arrange things so that this change to the $y_j$'s for $j\neq 0$ is sufficiently small that both inequalities in \eqref{eq:farkasdual} are satisfied, with the previously chosen value of $y_0$.

In detail this works as follows in this case. Towards the goal of finding a suitable $y$-vector, define the real function 
\begin{equation}
    f(x)=y_0 + 
y_1\cos(2\pi x)+
y_2\sin(2\pi x)
+y_3\cos(4\pi x)
+y_4\sin(4\pi x).
\end{equation}
Note that $y^T A\geq 0$ is equivalent to 
\begin{equation}
    f(x)\geq 0\mbox{ for all }x\in\{0,\ldots,N\}/2^k.
    \end{equation}
We will now exhibit values for the $y$-vector that satisfy a stronger condition than above, namely
\begin{equation}\label{eq:fpositivereal}
    f(x)\geq 0 \mbox{ for all }x\in[0,N/2^k].
    \end{equation}
    First, consider the case where $(y_0,y_1,y_2,y_3,y_4)=(0,-1/4, \sqrt{3}/{4}, 1/4,\sqrt{3}/{4})$. We will first analyse this choice of $y$ (with $y_0=0$) and show that the induced function~$f$ satisfies the inequality~\eqref{eq:fpositivereal}. Then,  following the strategy described in the previous paragraph but one,  we make $y_0$ negative in order to also satisfy the constraint $y^T E<0$, while adding a small function to $f$ (slightly changing the other $y_j$'s) in order to preserve the validity of \eqref{eq:fpositivereal} for an interval of $x$ whose right boundary is only slightly below $N/2^k$. 
    
\begin{figure}[htb]
    \centering
    \includegraphics[width=15cm]{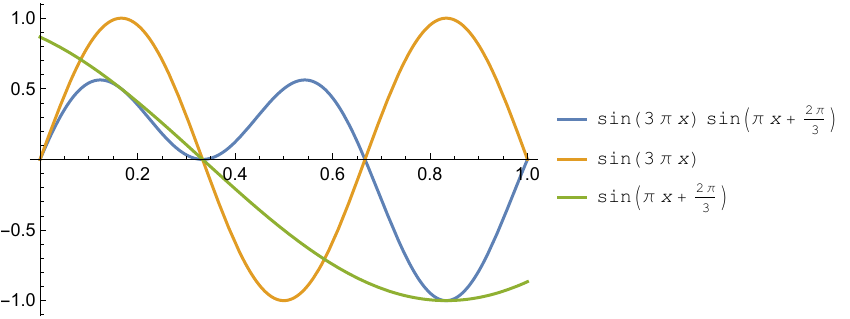}
\caption{The function $f_1(x)$ (in blue) and the two functions that make its product form}
\label{fig:f1}
\end{figure}

    Using trigonometric identities one may write the particular instantion of the function $f$ as follows (where the subscript in $f_1$ below indicates $\ell=1$, which will be generalized to larger $\ell$ in the next section)
\begin{equation}\label{eq:f1}
    f_1(x)= -\frac 1 4\cos(2\pi x)+
\frac{\sqrt{3}}{4}\sin(2\pi x)
+\frac 1 4\cos(4\pi x) +
\frac{\sqrt{3}}{4}\sin(4\pi x)=\sin(3\pi x)\sin(\pi x + \frac{2\pi}{3}).
\end{equation} 
Thus we may see from the right-hand side that $f_1(x)$ has zeros at 0, 1/3, and 2/3, and $f_1(x)$ is positive in between (since its two factors have the same sign between the zeros), satisfying Eq.~\eqref{eq:fpositivereal} for $N=\frac{2}{3}2^k$. This is illustrated in Figure~\ref{fig:f1}.

This $y\in\mathbb{R}^5$ does not necessarily satisfy the constraint $y^TE<0$ of Eq.~\eqref{eq:farkasdual} because $y_0=0$.
So we now choose $y_0$ to be sufficiently negative, setting it to $y_0=-2(|\eps_1|+|\eps_2|+|\eps_3|+|\eps_4|)-c(k)$, where $c(k)=o(1)$ is a strictly positive function. This implies 
\begin{equation}\label{eq:yTEbound}
    y^T E=y_0+\sum_{j=1}^4y_j\eps_j\leq -(|\eps_1|+|\eps_2|+|\eps_3|+|\eps_4|)-c(k)<0,
\end{equation}
 where we used that our chosen $y_1,y_2,y_3,y_4$ all have magnitude bounded below~1.

However the resulting function $f_1(x)+y_0$ has value $y_0<0$ at the zeros of $f_1$, so the condition of  Eq.~\eqref{eq:fpositivereal} doesn't hold anymore in an interval of width $o(1)$ around the zeros of $f_1$.  But we can modify the function by one final step to make it positive again at these points. We now add a small amount of the function 
$\sin(2\pi x+\pi/6)$. Specifically we form the function 
\begin{equation}
\tilde f_1(x)=f_1(x)+y_0 + 4|y_0|\sin(2\pi x+\pi/6).
\end{equation}
The key point is that $\sin(2\pi x+\pi/6)$ equals~$1/2$ at $x=0$ and $x=1/3$, and is at least $1/4$ in constant-sized intervals around $x=0$ and $x=1/3$. Hence adding $4|y_0|\sin(2\pi x+\pi/6)$ to $f_1(x)+y_0$ lifts the function above zero, compensating for the fact that $y_0$ is negative, and ensuring that the modified function $\tilde{f}_1$ is positive everywhere between $x=0$ and some point close to~$2/3$. The function $\sin(2\pi x+\pi/6)$ is  also expressible as a linear combination of $\cos(2\pi x)$ and $\sin(2\pi x)$, so we change $y_1$ and $y_2$ by an amount with magnitude $O(|y_0|)=o(1)$ to obtain $\tilde{f}_1$.
The inequality Eq.~\eqref{eq:yTEbound} remains valid because our modified $y_1$ and $y_2$ still have magnitude $\leq 1$,
so now we have satisfied both constraints of Eq.~\eqref{eq:farkasdual}.

This modification of $f_1$ to $\tilde{f}_1$ will reduce the value of $x$ near $x=2/3$ where the new function is zero (because $\sin(2\pi x+\pi/6)$ is negative around $x=2/3$), but only by an amount that can be made as small as we like, because $y_0=o(1)$. 
%
This proves that the linear system Eq.~\eqref{eq:farkasdual} is satisfiable for $N=\gamma\cdot 2^k$, where $\gamma$ can be made as close as we like to $2/3$; and therefore Eq.~\eqref{eq:farkasprimal} is \emph{un}satisfiable.
Hence we have proved a lower bound of $(\frac{2}{3}-o(1))2^k$ applications of $U_\phi$ that holds for every non-adaptive algorithm that can distinguish the 4 phases in the set $F$ with error probability~$o(1)$.

\section{A factor-2 advantage from larger phase-sets}\label{sec:2ellphases}

\subsection{The general problem and an upper bound of $\frac{1}{2}2^k$ for adaptive algorithms} 

Let us now consider a larger set of possible phases.
The previous set $F$ had 2 pairs of phases, one pair close to 0 at distance $2^{-k}$, and one pair close to 1/2 at distance $2\cdot 2^{-k}$.
We will now consider a set 
\begin{equation}
F_\ell \mbox{ with }2^\ell \mbox{ pairs of phases, one pair }r2^{-\ell},r2^{-\ell}+(r+1)2^{-k}\mbox{, for each }r=0,1,\ldots,2^{\ell}-1. 
\end{equation}
For example, the previous 4-phase set is $F_1$, and
$F_2$ is the 8-phase set 
\begin{equation}
\{0,2^{-k},~~1/4,1/4+2\cdot2^{-k},~~1/2,1/2+3\cdot 2^{-k},~~3/4,3/4+4\cdot2^{-k}\}
\end{equation}
Each phase $\phi\in F$ is a $k$-bit number whose leading  (i.e., most-significant) $\ell$ bits determine to which of the $2^\ell$ pairs the phase belongs; the last (i.e., least-significant) $\ell$ bits correspond to $(r+1)2^{-k}$; and in between those two $\ell$-bit strings is a sequence of $k-2\ell$ 0s.
Assuming $\ell$ is constant and $\ell\ll k$, there still is an adaptive algorithm for phase estimation with this phase set that uses $\frac{1}{2}2^k+O(2^\ell)\approx \frac{1}{2}2^k$ applications of $U_\phi$, as follows. First estimate the first $\ell$ bits of $\phi$ cheaply, using $O(2^\ell)$ applications  of $U_\phi$. This has error probability $O(2^{-k})$ thanks to the fact that there is a long sequence of at least $k-2\ell=k-O(1)$ 0s in $\phi$ following its first $\ell$ bits. Learning the $\ell$ most significant-bits narrows down the possible phases to one pair $r2^{-\ell},r2^{-\ell}+(r+1)2^{-k}$. Second, use at most $\frac{1}{2}2^k$ applications of~$U_\phi$ to learn (with success probability~1) the least-significant bit of~$\phi$. That bit is~1 for the phase $r2^{-\ell}+(r+1)2^{-k}$, and 0 for the phase $r2^{-\ell}$, so learning that one bit suffices to distinguish the two phases in the remaining pair.

\subsection{An upper bound of $\frac{2^\ell}{2^\ell+1}2^k$ for non-adaptive algorithms}

We claim that there is a non-adaptive algorithm with essentially $\frac{2^\ell}{2^\ell+1}2^k$ applications of $U_\phi$ to distinguish these phases. This is essentially optimal for $\ell=1$ as we showed above (via Farkas's lemma), and for $\ell\geq 2$ as we will show below in Section~\ref{sec:ellLB}.

First, using $2^{\ell}-1$ applications of $U_\phi$ we can learn the $\ell$ most-significant bits of $\phi$ by standard phase estimation ``flattened out'' on many qubits to make it non-adaptive (as explained in Section~\ref{ssec:basicsensing}).
Once we know which multiple $r2^{-\ell}$ of $2^{-\ell}$ our unknown phase $\phi$ is close (or equal) to, it remains to distinguish the two phases $r2^{-\ell}$ and $r2^{-\ell}+(r+1)2^{-k}$, i.e., to resolve a phase difference $(r+1)2^{-k}$. The crucial thing that distinguishes non-adaptive from adaptive algorithms here, is that a non-adaptive algorithm cannot use a state that depends on $r$: it has to use the same $\beta$-vector of amplitudes for all $r$.

 Let $N$ be $\frac{2^\ell}{2^\ell+1}2^k$, rounded to the nearest multiple of $2^{2\ell}$. The rounding induces a shift $\Delta=N-\frac{2^\ell}{2^\ell+1}2^k$ of magnitude $|\Delta|<2^{2\ell-1}$ compared to the intended number $\frac{2^\ell}{2^\ell+1}2^k$; it ensures that $N/2^\ell$ is an integer multiple of $2^\ell$. 
 Now consider the following state, which is supported on integer multiples of $N/2^\ell$:
\begin{equation}
    \frac{1}{\sqrt{2^\ell+1}}\sum_{m=0}^{2^\ell}\ket{\frac{mN}{2^\ell}}.
\end{equation}
The largest integer in this superposition is indeed $N$, which is $\frac{2^\ell}{2^\ell+1}2^k+\Delta$, and which corresponds to the number of applications of $U_\phi$ of the non-adaptive algorithm. 
Note that the state of Eq.~\eqref{eq:stateforell=1} is the $\ell=1$ case, where $N$ is $\frac{2}{3}2^k$ rounded to the nearest multiple of~4.

If we apply $\ket{j}\mapsto e^{2\pi i j \phi}\ket{j}$,  then we obtain the state
\begin{equation}\label{eq:nonadapstates} \frac{1}{\sqrt{2^\ell+1}}\sum_{m=0}^{2^\ell}e^{2\pi i mN\phi/2^\ell}\ket{\frac{mN}{2^\ell}},
\end{equation}
Note that because $N/2^\ell$ is an integer multiple of $2^\ell$, the phases in this superposition do not depend on the first $\ell$ bits of $\phi$. To show that this non-adaptive algorithm works, it now suffices to show near-orthogonality  for the states induced by phases $0$ and $(r+1)2^{-k}$, for each $r=0,\ldots,2^\ell-1$, because then there exists a measurement that distinguishes all $|F|$ states with exponentially small error probability (using the first part of the state to learn the first $\ell$ bits of $\phi$ and the state of \eqref{eq:nonadapstates} to resolve the difference between the pair of phases induced by the first $\ell$ bits). 
The inner product between the two states for phases $(r+1)2^{-k}$ and~$0$ is
\begin{equation}\label{eq:iptwophases}
\frac{1}{2^\ell+1}\sum_{m=0}^{2^\ell}e^{2\pi i mN(r+1)2^{-k}/2^\ell}
=\frac{1}{2^\ell+1}\sum_{m=0}^{2^\ell}e^{2\pi i m((r+1)/(2^{\ell}+1)+\Delta((r+1)2^{-k})/2^\ell)},
\end{equation}
where the equality used that 
$N=\frac{2^\ell}{2^\ell+1}2^k+\Delta$.
We have $m\Delta((r+1)2^{-k})/2^\ell\leq 2^{3\ell-1-k}$ (because $m\leq 2^\ell,\Delta\leq 2^{2\ell-1},r\leq 2^\ell-1$), 
which is an exponentially small (in $k$)  change to the phase $e^{2\pi i m(r+1)/(2^{\ell}+1)}$  because we think of $\ell$ as a constant and $k$ growing.
Without that small change in the phase, the geometric sum
$\sum_{m=0}^{2^\ell}e^{2\pi i m(r+1)/(2^{\ell}+1)}=0$.
Hence the inner product of \eqref{eq:iptwophases} is exponentially small (in $k$).

\subsection{A lower bound of $\frac{2^\ell}{2^\ell+1}2^k$ for non-adaptive algorithms}\label{sec:ellLB}

We now extend the argument for the lower bound of Section~\ref{ssec:LBell=1} to $\ell\ge 2$.
We again give a lower bound via Farkas's lemma, by providing a function that is positive for $x\in [0,\frac{2^\ell}{2^\ell+1}]$ and that is a sum of sines and cosines of the form
\begin{equation}
\sum_{j=1}^{2^\ell} \Big(y_{2j-1} \cos(2\pi j x) + y_{2j} \sin(2\pi j x)\Big),\label{eq:sum-form}
\end{equation}
where the coefficients $y_1,\ldots,y_{2^{\ell+1}}$ are real.

Here is the particular  function we will consider
\begin{equation}
f_\ell(x)=\sin \left(\pi  \left(x-\frac{1}{2^\ell+1}\right)\right) \sin \left(\pi  \left(2^\ell+1\right) x\right)\prod_{m=2}^{2^{\ell-1}} \Big( \cos(2 \pi\frac{  m}{2^\ell + 1}) -  \cos(2 \pi x)\Big).\label{eq:product-form}
\end{equation}

\begin{figure}[htb]
    \centering
    \includegraphics[width=10cm]{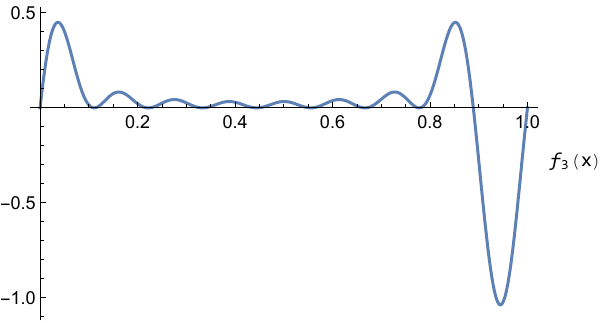}
    \caption{$f_3(x)$ as defined in \eqref{eq:f3}
    }
    \label{fig:f_3}
\end{figure}

\begin{figure}[htb]
    \centering
    \includegraphics[width=15cm]{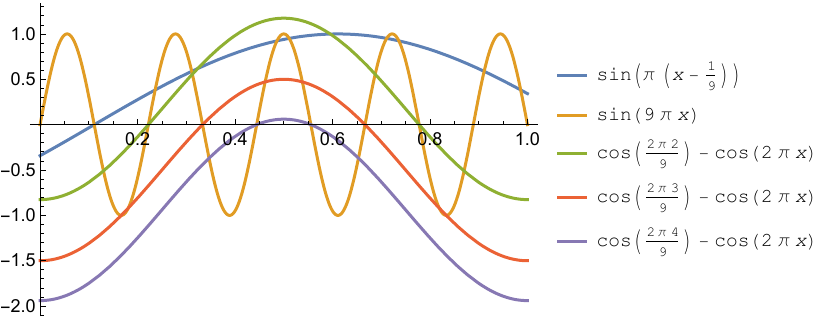}
    \caption{The five functions in the product form for $f_3(x)$}
    \label{fig:f_3-components}
\end{figure}

We will show that this function is positive for $x\in [0,\frac{2^\ell}{2^\ell+1}]$, and, in the region $x\in [0,1]$ only has zeros at integer multiples of $1/(2^\ell+1)$. To understand why this function satisfies our desiderata, it will be instructive to first consider a particular case, for example $\ell=3$:
\begin{equation}
f_3(x)=\sin \big(\pi  (x-\frac{1}{9})\big) \sin \left(9\pi  x\right)\prod_{m=2}^{4} \Big( \cos(2 \pi\frac{  m}{9}) -  \cos(2 \pi x) \Big).\label{eq:f3}
\end{equation}
The function $f_3(x)$ together with the five component functions in the product form are illustrated in Figures \ref{fig:f_3} and \ref{fig:f_3-components}, respectively.

We are only interested in the region $x\in [0,1]$, so in what follows, any statements only concern values $x$ in this region.
In order to understand why $f_3(x)$ is positive on $[0,8/9]$, it is helpful to consider the signs of the component functions in the 9 regions $[1,1/9], [1/9,2/9],\ldots, [8/9,1]$. 

Firstly, we note that $\sin(9\pi x)$ alternates in sign in each successive region of width $1/9$: its sign pattern is 
$(+,-,+,-,+,-,+,-,+)$.  

Then consider the function
\begin{equation}
    g_m(x)=\cos(2 \pi\frac{  m}{9}) -  \cos(2 \pi x) 
\end{equation}
for $m=2,3,4$.  In the region $x\in [0,1]$, $g_m(x)$ only has zeros at $x=m/9$ and $x=1-m/9$, and it is positive for $x\in [m/9,1-m/9]$ and negative otherwise.  Thus, if we consider each successive region of length $1/9$, the functions $g_2,g_3,g_4$ have sign patterns, 
$(-,-,+,+,+,+,+,-,-)$, $(-,-,-,+,+,+,-,-,-)$, $(-,-,-,-,+,-,-,-,-)$, respectively. 

Thus the product 
\begin{equation}
\prod_{m=2}^{4} \Big( \cos(2 \pi\frac{  m}{9}) -  \cos(2 \pi x) \Big)
\end{equation}
has sign pattern $(-,-,+,-,+,-,+,-,-)$, and therefore the product 
\begin{equation}
\sin \left(9\pi  x\right)\prod_{m=2}^{4} \Big( \cos(2 \pi\frac{  m}{9}) -  \cos(2 \pi x) \Big)
\end{equation}
has sign pattern $(-,+,+,+,+,+,+,+,-)$. Lastly, $\sin \big(\pi  (x-\frac{1}{9})\big)$ is negative for $x\in [0,1/9]$ and positive otherwise, so its sign pattern is $(-,+,+,+,+,+,+,+,+)$. Accordingly, for each of the first 8 regions, multiplying out the signs of the five component functions gives $+$, so $f_3(x)$ is positive in $x\in [0,8/9]$, as required.

Furthermore (for example by writing each cosine and sine using $\cos a= (e^{ia}+e^{-ia})/2$ and $\sin a= (e^{ia}-e^{-ia})/2i$)  it may be seen that $f_3(x)$ only contains the required cosines and sines as in Eq.~\eqref{eq:sum-form}.  

Just as in the case $\ell=1$ in Section~\ref{ssec:LBell=1}, we need to modify $f_3$ to make sure both parts of the requirements of Farkas's lemma (Eq.~\eqref{eq:farkasdual}) are satisfied.  
In particular, the constraint $y^T E<0$ is not guaranteed  to hold for the choice of $y_j$'s (incl.~$y_0=0$) that define the current~$f_3$. We again make two modifications.  Firstly we add a small negative $y_0$;  namely let $Y_{\rm max}$ be the maximum value of the modulus of the $y_j$, for $j\neq 0$.\footnote{We note that a simple upper bound for  $f_3$ is $2^3$. This upper bounds (using triangle inequality) the $y_j$ since they are the Fourier coefficients of $f_3$ when decomposing $f_3$ as a linear combination of sines and cosines; hence in fact $Y_{\rm max}\leq 2^4$, but all that is needed for our purposes is that it is upper bounded by a constant as $k\rightarrow \infty$.} 
Then we take
\begin{equation}\label{eq:y0defn}
    y_0=-2Y_{\rm max}\sum_{j=1}^{2^{\ell+1}} |\epsilon_j|.
    \end{equation} 
Secondly, the addition of this negative $y_0$ means the function $f_3+y_0$ will be negative around the zeros of $f_3$, so we need to add a function to $f_3+y_0$ to make sure we get a function that is positive in the entire required region.  Fortunately, there is a simple choice at hand: using similar arguments to the ones we have just made for~$f_3$, the function $f_2$ is strictly positive at the places where $f_3$ is zero in the region $x\in [0,8/9)$. (The zeros of $f_2$ are at integer multiples of $1/5$, so they do not coincide with the zeros of $f_3$; and $f_2$ is positive for $x\in [0,4/5]$.)  Thus we can add a small multiple of $f_2$ --- shifted slightly to the left (by $1/90$, which is half the minimal distance between the zeros of $f_3$ and the zeros of $f_2$) --- to compensate for the addition of the small negative $y_0$.  This shifted function $f_2$ will be strictly positive at the zeros of $f_3$. Let $z_{\rm min}$ denote the minimum value of the shifted $f_2$ at these zeros. 
Because $f_2$ has bounded derivative, it will then have value at least $z_{\rm min}/2$ in constant-width intervals around the zeros of $f_3$.

Thus our modified function 
\begin{equation}
\tilde f_3(x)=f_3(x)+y_0+\frac{2|y_0|}{z_{\rm min}}f_2(x+1/90)
 \end{equation}
 will be positive from $x=0$ almost up to $x=8/9$, and contain only the allowed sines and cosines. This ensures that he first inequality in Farkas's lemma ($y^T A \geq 0$) holds for our new coefficients~$y_j$.
 The coefficients $y_j$ for $\tilde f_3$ are slightly different from those for $f_3$; but since $|y_0|=o(1)$ (i.e., $y_0$ goes to 0 as $k\rightarrow \infty$), and $z_{\rm min}$ and all the $y_j$ for $f_2$ are independent of $k$,  the second inequality in Farkas's lemma ($y^T E< 0$) will still hold.
This was the reason for having the factor~2 in Eq.~\eqref{eq:y0defn}. 

As in the case of $\ell=1$ in Section~\ref{ssec:LBell=1}, this addition of $y_0$ and $f_2$ will slightly reduce the region of $x$ for which the function is positive, but only by an amount that we can make as small as we like by starting with a sufficiently small error probability~$\errorprob$, since the changes to the $y_j$ are $O(\sqrt\errorprob)$ and $\kappa=o(1)$.


The general argument about the positivity of $f_\ell(x)$ for $\ell$ other than~3, proceeds in the same way. As before we are only interested in the region $x\in [0,1]$. The factors after the product symbol in Eq.~\eqref{eq:product-form} are positive in the region $x\in [m/(2^\ell+1),1-m/(2^\ell+1)]$ and negative otherwise. So we consider regions of $x$ of width $1/(2^\ell+1)$. The product of these functions has the same pattern of signs as $\sin \left(\pi  \left(2^\ell+1\right) x\right)$, apart from in the first region $x\in [0,1/(2^\ell+1)]$, but the remaining factor in Eq.~\eqref{eq:product-form}, namely $\sin \left(\pi\left(x-\frac{1}{2^\ell+1}\right)\right)$, is negative in $x\in [0,1/(2^\ell+1)]$, and positive otherwise, so that $f_\ell(x)$ is indeed positive for all $x\in [0,2^\ell/(2^\ell+1)]$.  And $f_\ell(x) $ only contains the required cosines and sines as in Eq.~\eqref{eq:sum-form}.  Finally, as in the discussion for $\ell=3$, we can arrange both requirements of Farkas's lemma to be satisfied by adding a small negative $y_0$ and by adding a small amount of a shifted version of $f_{\ell-1}$ to $f_\ell$, noting that  $f_{\ell-1} $ and $f_\ell$ have no common zeros in $[0,1)$\footnote{Such a common zero in $[0,1)$ would have to be an integer multiple of $1/(2^\ell+1)$ and of $1/(2^{\ell-1}+1)$, but that doesn't exist because the numbers $2^{\ell}+1$ and $2^{\ell-1}+1$ are coprime (because if $p>1$ is a factor of $2^{\ell-1}+1$, then $(2^{\ell}+1)/p=2(2^{\ell-1}+1)/p-1/p$ is an integer minus $1/p$, which is not an integer).}, and the region in which $f_{\ell-1}$ is positive extends beyond $x\in [0,(2^\ell-1)/(2^\ell+1)]$.\footnote{The size of $y_0$ and the ``small amount'' of $f_{\ell-1} $ could grow very rapidly with $\ell$ but this is not an issue since we think of $\ell$ as a constant and $k\rightarrow\infty$.}

\section{Are bigger separations possible?}\label{sec:bettersep?}

Our examples naturally lead to the question whether there could be bigger separations between the number of applications of~$U_\phi$ in adaptive and non-adaptive phase estimation, significantly bigger than our factor of~2.

If all phases in $F$ can be written as $k$-bit numbers, and the smallest difference $\delta=\min\{|\phi-\phi'| \mbox{ mod}~1 \mid \mbox{distinct }\phi,\phi'\in F\}$ between any two phases in $F$ is $\delta=2^{-k}$ (as it is in the specific sets $F_\ell$ that we use for our separations), then we can show there is no separation larger than factor-2, as follows. On the one hand, the parallelized version of standard $k$-bit phase estimation non-adaptively  recovers the phase~$\phi$ \emph{with success probability~1} using $2^k-1<1/\delta$ applications of~$U_\phi$. But on the other hand, even just to resolve the difference between the minimizing phases $\phi$ and $\phi'$, an adaptive phase estimation algorithm needs at least $(1/2-o(1))/\delta$ applications of~$U_\phi$ if we want to have error probability $o(1)$, by the lower bound argument of  Section~\ref{subsection:our-result}. 
So for large constant~$\ell$ our separations based on the sets $F_\ell$ approach the maximal separation (i.e., factor-2) that is possible for sets $F$ where $k$ bits of precision are necessary and sufficient.

What about phase sets~$F$ where $\delta\approx 2^{-k}$ but some of the phases $\phi\in F$ need more than $k$ bits of precision?
The lower bound of $(1/2-o(1))/\delta\approx \frac{1}{2}2^k$ applications of~$U_\phi$ for adaptive algorithms that have error probability $o(1)$, remains valid.
Non-adaptively, we can still do parallelized phase estimation with $k$ bits of precision. 
This will use $2^k-1$ applications of $U_\phi$, and will have some \emph{constant} error probability but not necessarily error probability $o(1)$. Hence we cannot conclude now that factor-2 separation is best possible when comparing adaptive and non-adaptive algorithms that have vanishing error probability.
In general one can show that the standard phase estimation algorithm with $t$ bits of precision has additive error $\leq 2^{-k}$ in its estimate, except with error probability $O(2^{k-t})$, see for instance \cite[Eq.~(5.35)]{nielsen&chuang:qc}.
We can ensure that the non-adaptive phase estimation has error probability $o(1)$ by using $t=k+b(k)$ bits of precision, for any function $b(k)$ that goes to infinity, even extremely slowly.\footnote{For instance we could use $b(k)=\log^*(k)$, which is defined as the number of times the  binary logarithm needs to be applied to $k$ iteratively before the value drops below~1.} 
However, the number of applications of~$U_\phi$ of this non-adaptive phase estimation now would be $2^{k+b(k)}-1$, and the ratio with the adaptive lower bound of roughly $\frac{1}{2}2^k$ would go to infinity with~$k$, albeit arbitrarily slowly.
It is tempting to conjecture that there exists some universal constant $C$ (possibly even $C=2$) such that the adaptive advantage is never more than factor-$C$ for any phase-set~$F$. However, we were not able to prove this and leave it as a question for future work. 


\section{Conclusion and discussion}

We studied phase estimation on a single-qubit phase-gate $U_\phi$ (which is a basic scenario in quantum sensing) in the setting where we have a promise on the values that the unknown phase~$\phi$ can take. This twist on the phase-estimation problem is possibly of some practical interest in situations where we have some prior information about the value of~$\phi$. We provided examples where adaptive algorithms had  an advantage of nearly a factor of~2 over non-adaptive algorithms in terms of the number of applications of $U_\phi$ needed---in contrast to the promiseless scenario, where there is no such advantage. However we also proved that such an advantage cannot be very large: at most a factor of 2 when $k$ bits of precision are necessary and sufficient for the promised phases, and at most an arbitrarily-slowly growing function in general (it would be very interesting to close that gap). We introduced some new techniques here that may find use elsewhere, for instance the use of Farkas's lemma to lower bound the required number of non-adaptive uses of $U_\phi$: disproving the existence of a suitable initial amplitude-vector by exhibiting a solution to a dual problem.

As mentioned in the introduction, one of the main motivations for studying fully parallel ($=$ non-adaptive) algorithms is that they are less sensitive to noise on the qubits due to their lower depth. Hence a next step would be to study phase estimation in the presence of various kinds of noise, and to study to what extent quantum advantages (both adaptive and non-adaptive) disappear.
There is already a significant literature studying in what situations the desired Heisenberg scaling (where $O(1/\delta)$ applications of $U_\phi$ suffice to estimate $\phi\pm\delta$) gives way to so-called ``classical'' scaling (where $\Omega(1/\delta^2)$ applications of $U_\phi$ are needed), see for instance~\cite{DKM:elusiveheisenberg}.
One could similarly ask to what extent adaptive-over-non-adaptive advantages survive in the presence of various kinds of noise.

%
%

\paragraph{Acknowledgments.}
Mathematica~\cite{Mathematica} was used for the figures in Sections~\ref{ssec:LBell=1} and~\ref{sec:ellLB}, and for algebraic experiments earlier in the work. We thank
 Ashley Montanaro for providing us with Figure~\ref{fig:PE}.


\bibliographystyle{alpha}
\bibliography{qc}

\appendix

\section{Maximum angle that controlled-$U_\phi$ can induce}

Given two quantum states $\ket a$ and $\ket b$ we define the angle $\theta\in[0,\pi/2]$ between them by
\begin{equation}|\langle a \ket b| = \cos \theta.
\end{equation}
Let $\ket\Psi$ be a normalised state of $2+q$ qubits. We can write
\begin{equation}\ket\Psi = \ket{00}\ket{\psi_{00}}+\ket{01}\ket{\psi_{01}}+\ket{10}\ket{\psi_{10}}+\ket{11}\ket{\psi_{11}}.
\end{equation}
So 
\begin{eqnarray}
C(U_\phi) \otimes I \ket\Psi &=& \ket{00}\ket{\psi_{00}}+\ket{01}\ket{\psi_{01}}+\ket{10}\ket{\psi_{10}}+e^{2\pi i \phi}\ket{11}\ket{\psi_{11}}\nonumber\\
&:=& \ket\eta +e^{2\pi i \phi}\ket{11}\ket{\psi_{11}}.
\end{eqnarray}
We want the maximum angle that the controlled $U_\phi$ can induce, which is equivalent to minimizing the inner product between $\ket\Psi$ and $C(U_\phi) \otimes I \ket\Psi$ 
\begin{eqnarray}
{\rm min}_{\ket\Psi} \left|\bra\Psi C(U_\phi) \otimes I \ket\Psi\right|^2 &=& 
 {\rm min}_{\ket\Psi} \left|\langle \eta\ket \eta + e^{2\pi i \phi}\langle\psi_{11}\ket{\psi_{11}}\right|^2\nonumber\\
 &=&  \min_{z\in[0,1]}  \left|1-z + z e^{2\pi i \phi}\right|^2,
\end{eqnarray}
where we have defined $\langle\psi_{11}\ket{\psi_{11}}=z$.
Thus 
\begin{eqnarray}
\min_{\ket\Psi} \left|\bra\Psi C(U_\phi) \otimes I \ket\Psi\right|^2 
 &=&   \min_{z\in[0,1]}  \quad 1 + 2z(1-z)(\cos(2\pi\phi)-1)\nonumber\\
 &=& 1+ \frac 1 2(\cos(2\pi\phi)-1) = \cos^2(\pi\phi) .
\end{eqnarray}
Hence the maximum angle induced is no more than $\pi\phi$.

\section{Triangle inequality for angles between states}

Consider three states $\ket a,\ket b, \ket c$, with inner products
\begin{equation}
\langle a \ket b = e^{i\theta_1}\cos \alpha,\quad 
\langle b \ket c = e^{i\theta_2}\cos \beta,\quad
\langle c \ket a = e^{i\theta_3}\cos \gamma.\quad
\end{equation}
We choose $0\leq \alpha, \beta, \gamma \leq \pi/2$.  We will show that $\gamma\leq \alpha + \beta $.

We note that we could choose phases for $\ket a,\ket b, \ket c$ to make the phases of two of the inner products zero, but not all three.%
\footnote{For example by noting that $\langle a \ket b\langle b \ket c \langle c \ket a = \Tr(\ \ket a\bra a\cdot  \ket b\bra b\cdot \ket c\bra c\ )$ is invariant under choice of phases.}

The Gram matrix\footnote{We have adapted the argument that is given for real vectors in \url{https://math.stackexchange.com/questions/4670434/does-angle-between-vectors-satisfy-triangle-inequality}} $G_{jk}=\langle \psi_j\ket {\psi_k}$ of any set of states $\{\ket{\psi_j}\}$ is positive semi-definite, so in particular its determinant is $\ge 0$.  The determinant of the Gram matrix of the states $\ket a,\ket b, \ket c$ is
\begin{equation} 1-\cos^2\alpha -\cos^2\beta -\cos^2\gamma + 2 \cos\alpha \cos\beta\cos \gamma \cos(\theta_1+\theta_2+\theta_3).\end{equation}
Thus 
\begin{equation} 1-\cos^2\alpha -\cos^2\beta -\cos^2\gamma + 2 \cos\alpha \cos\beta\cos \gamma \ge 0 .\end{equation}
But 
\begin{equation} 1-\cos^2\alpha -\cos^2\beta -\cos^2\gamma + 2 \cos\alpha \cos\beta\cos \gamma = 4 \sin(p)\sin(p-\alpha)\sin(p-\beta)\sin(p-\gamma),\label{Gram} \end{equation}
where $p=\frac 1 2 (\alpha +\beta +\gamma)$.

Now $\sin p \ge 0$.  Without loss of generality, let us take the angles ordered as $\alpha \leq \beta \leq \gamma$. In order for the right-hand side of \eqref{Gram} to be positive, either (a) $p\ge \alpha$ and $p\le \beta, \gamma$, or (b) $p\ge \gamma$ (and hence $p\ge \alpha, \beta$).  In the former case, however $p\le \beta, \gamma$ implies that $\alpha\leq 0$, contrary to our assumptions about $\alpha$ (other than the trivial case of $\alpha=0$).  Thus we are left with case (b), i.e. $(\alpha+\beta+\gamma)/2\ge \gamma,$ in other words $\alpha + \beta \ge \gamma$ as required.
\end{document}